\documentclass[useAMS,usenatbib]{mn2e}
\usepackage{mncite}
\usepackage{graphicx}
\usepackage{hyperref}
\usepackage{amssymb}
\usepackage{aas_macros}
\usepackage{subfig}
\usepackage[para]{threeparttable}
\usepackage{color}

\title[Are falling planets spinning up their host stars?]{Are falling planets spinning up their host stars?}
\author[D. J. A. Brown et al]
{
D. J. A. Brown$^{1}$\thanks{E-mail:djab@st-andrews.ac.uk},
A. Collier Cameron$^{1}$,
C. Hall$^{1}$,
L. Hebb$^{1}$,
B. Smalley$^{2}$,
\\
$^{1}$SUPA, School of Physics and Astronomy, University of St Andrews, North Haugh, St Andrews, Fife KY16 9SS, UK.\\
$^{2}$Astrophysics Group, School of Chemistry and Physics, Keele University, Staffordshire, ST5 5BG, UK.\\
}
\begin{document}

\date{Accepted 0000 December 00. Received 0000 December 00; in original form 0000 October 00}

\pagerange{\pageref{firstpage}--\pageref{lastpage}} \pubyear{2008}

\maketitle

\label{firstpage}

\begin{abstract}
We investigate the effects of tidal interactions on the planetary orbits and stellar spin rates of the WASP-18 and WASP-19 planetary systems using a forward integration scheme. By fitting the resulting evolutionary tracks to the observed eccentricity, semi-major axis and stellar rotation rate, and to the stellar age derived from isochronal fitting, we are able to place constraints on the stellar and planetary reduced tidal quality factors, $Q'_s$ and $Q'_p$. We find that for WASP-18, $\log(Q'_s)=8.21^{+0.90}_{-0.52}$ and $\log(Q'_p)=7.77^{+1.54}_{-1.25}$, implying a system age of $0.579^{+0.305}_{-0.250}$\,Gyr. For WASP-19 we obtain values of $\log(Q'_s)=6.47^{+2.19}_{-0.95}$ and $\log(Q'_p)=6.75^{+1.86}_{-1.77}$, suggesting a system age of $1.60^{+2.84}_{-0.79}$\,Gyr and a remaining lifetime of $0.0067^{+1.1073}_{-0.0061}$\,Gyr. We investigate a range of evolutionary histories consistent with these results and the observed parameters for both systems, and find that the majority imply that the stars have been spun up through tidal interactions as the planets spiral towards their Roche limits. We examine a variety of evidence for WASP-19\,A's age, both for the value above and for a younger age consistent with gyrochronology, and conclude that the older estimate is more likely to be correct. This suggests that WASP-19\,b might be in the final stages of the spiral-in process, although we are unable to rule out the possibility that it has a substantial remaining lifetime.
\end{abstract}


\section{Introduction}
There are currently 538 officially confirmed extrasolar planets known to astronomy\footnote{As of 2011 March 15. A full list of known exoplanets can be found at http://exoplanet.eu}, a substantial fraction of which belong to the `hot Jupiters' category of Jovian-type planets that orbit within $0.1$\,AU of their host star. Improved observational precision has, in recent years, led to the discovery of extreme examples of this class of planet with orbital semi-major axes of the order of $0.02$\,AU or less(see, for example, \citet{sasselov2003}, \citet{hellier2009}, \citet{hebb2010}). The Jovian masses and small orbital separations of these planets imply that they interact tidally with their host stars.

Tidal interactions between stars and planets were first studied in the context of the Solar system \citep{goldreich1966}, and have also been extensively studied as they apply to binary systems \citep{hut1980}, but it is only comparatively recently that such studies have been extended to consider exoplanetary systems. Tidal interactions lead to long-term changes in the orbital parameters of a planet, specifically the eccentricity and the semi-major axis (see e.g. \citet{mardling2002}; \citet{jackson2008}; \citet{barker2009}; \citet{jackson2009}). The end result of this process is the spiral-in of the planet towards its host until it reaches the Roche limit, where it undergoes mass transfer through the L1 point \citep{gu2003}, becoming disrupted in the process. Previous studies of tidal evolution have, for the most part, focused on the effect of tidal interactions on these two orbital parameters whilst neglecting the evolution of other parameters that are involved. 

\citet{dobbsdixon2004} demonstrated that the effects of tidal interactions extend to the rotation rates of the two bodies, but until recently little work had been done to investigate this aspect of tidal theory as applied to exoplanetary systems. There is, however, mounting interest in this aspect of exoplanetary tides, with several studies investigating the large scale effect of the presence of exoplanets on stellar rotation \citep{pont2009,alves2010,lanza2010}, and some work investigating tidal interactions specifically \citep{leconte2010}. It is well known that the rotation rate of an isolated star declines with age through the action of magnetic braking \citep{weberdavis1967}, which leads asymptotically to a power-law decay of rotation rate with the inverse square root of stellar age \citep{skumanich1972}. \citet{barnes2007} presented a method for determining the ages of stars using their rotation periods and colours, based on a Skumanich-type magnetic braking law and calibrated using the solar rotation period, as an alternative to stellar model fitting. What appears to be less widely recognized, or at least acknowledged, is that the torques that act on both star and planet as a result of tidal interactions can affect this natural rotational evolution, and may be sufficiently strong to overwhelm it, at least for short periods of time.

In this paper we simulate the evolution of hot Jupiter systems to investigate the effects on the orbital eccentricity, orbital separation, planetary rotation rate and stellar rotation rate of tidal interactions between hot Jupiters and their host stars. In Section\,\ref{sec:maths} we set out the mathematical basis for our models, whilst in Section\,\ref{sec:compute} we describe the computational methods that were used. Section\,\ref{sec:wasp18} and Section\,\ref{sec:wasp19} contain discussion of the application of our simulations to the specific cases of the WASP-18 and WASP-19 systems respectively. It should be noted that these systems were not selected at random, but were chosen owing to pre-existing disagreement between age estimates calculated using different methods. These transiting hot Jupiters also have two of the shortest known orbital periods, and are therefore particularly susceptible to the influence of tidal interactions.

\section{Tidal and wind evolution}
\label{sec:maths}

Following \citet{eggleton1998}, \citet{mardling2002} and \citet{dobbsdixon2004} tidal energy is dissipated within a star and planet whose spin axes are aligned with the orbital axis at rates defined by the tidal quality factors $Q'_s=3Q_s/2k_*$ and $Q'_p=3Q_p/2k_p$, where $k_*$ and $k_p$ are the tidal Love numbers of the two bodies. The eccentricity of the orbit evolves at a rate
\begin{eqnarray}
\frac{\dot{e}}{e}&=&\frac{81}{2}\frac{n}{Q'_p}\frac{M_s}{M_p}\left(\frac{R_p}{a}\right)^5
\left[-f1(e)+\frac{11}{18}\frac{\Omega_p}{n}f_2(e)\right]+\nonumber\\
&&+\frac{81}{2}\frac{n}{Q'_s}\frac{M_p}{M_s}\left(\frac{R_s}{a}\right)^5
\left[-f1(e)+\frac{11}{18}\frac{\Omega_s}{n}f_2(e)\right]
\label{eq:dlnedt}
\end{eqnarray}
where
\begin{equation}
f_1(e)=\left(1+\frac{15}{4}e^2+\frac{15}{8}e^4+\frac{5}{64}e^6\right)/(1-e^2)^{13/2}
\label{eq:f1}
\end{equation}
and
\begin{equation}
f_2(e)=\left(1+\frac{3}{2}e^2+\frac{1}{8}e^4\right)/(1-e^2)^5.
\label{eq:f2}
\end{equation}
The star and planet have masses $M_s$ and $M_p$ and radii $R_s$ and $R_p$ respectively, and rotation rates $\Omega_s$ and $\Omega_p$. The orbital frequency is defined by Kepler's 3rd law, $n^2=G(M_s+M_p)/a^3$.

If the system is fully aligned, then the total angular momentum of the orbit and the axial rotation of the star and the planet, perpendicular to the orbit, is
\begin{equation}
L_{\rm tot} = M_p M_s\sqrt{\frac{Ga(1-e^2)}{M_p+M_s}}+\alpha_sM_sR_s^2\Omega_s+\alpha_pM_pR_p^2\Omega_p.
\label{eq:ltot}
\end{equation}
The stellar and planetary moments of inertia are determined by their effective squared radii of gyration, $\alpha_s$ and $\alpha_p$.

Angular momentum is carried away from the system via a magnetically-channeled, thermally-driven stellar wind, at a rate described by a standard Weber-Davis model
\begin{equation}
\dot{L}_{\rm wind} = -I_s\kappa\Omega_s{\rm Min}(\Omega_s,\tilde{\Omega})^2
\label{eq:ldotwind}
\end{equation}
where the stellar moment of inertia $I_s=\alpha_sM_sR_s^2$. The physical scaling of the braking rate is determined by the constant of proportionality $\kappa$, and $\tilde{\Omega}$ is the `saturation' rotation rate above which the stellar magnetic field strength is assumed to become independent of the stellar rotation rate.

We use the expressions of \citet{dobbsdixon2004} to describe the tidal spin evolution of the planet to the tidal torque:
\begin{equation}
\dot{\Omega}_p =
\frac{9}{2}
\left(\frac{n^2}{\alpha_pQ'_p}\right)
\left(\frac{M_s}{M_p}\right)
\left(\frac{R_p}{a}\right)^3
\left[f_3(e)-f_4(e)\frac{\Omega_p}{n}\right]
\label{eq:dompdt}
\end{equation}
and the star under the influence of both tidal and wind torques:
\begin{eqnarray}
\dot{\Omega}_s &= &
\frac{9}{2}
\left(\frac{n^2}{\alpha_sQ'_s}\right)
\left(\frac{M_p}{M_s}\right)^2
\left(\frac{R_s}{a}\right)^3
\left[f_3(e)-f_4(e)\frac{\Omega_s}{n}\right]
\nonumber\\
&&
-\kappa\Omega_s{\rm Min}(\Omega_s,\tilde{\Omega})^2.
\label{eq:domsdt}
\end{eqnarray}
The polynomials $f_3(e)$ and $f_4(e)$ have the form 
\begin{equation}
f_3(e)=\left(1+\frac{15}{2}e^2+\frac{45}{8}e^4+\frac{5}{16}e^6\right)/(1-e^2)^6
\label{eq:f3}
\end{equation}
and
\begin{equation}
f_4(e)=\left(1+3e^2+\frac{3}{8}e^4\right)/(1-e^2)^{9/2}.
\label{eq:f4}
\end{equation}

The total angular momentum of the system evolves as
\begin{eqnarray}
\dot{L}_{\rm wind} &=& M_p M_s\sqrt{\frac{G}{M_p+M_s}}
\left[
\frac{\dot{a}(1-e^2)-2ae\dot{e}}{2\sqrt{a(1-e^2)}}
\right]+\nonumber\\
&&+\alpha_sM_sR_s^2\dot{\Omega}_s+\alpha_pM_pR_p^2\dot{\Omega}_p.
\label{eq:ldottotwind}
\end{eqnarray}

Dividing the angular momentum loss rate by the orbital angular momentum 
\begin{equation}
L_{\rm orb}=M_p M_s\sqrt{\frac{Ga(1-e^2)}{M_p+M_s}}
\label{eq{lorb}}
\end{equation}
and rearranging, we obtain the expression for the evolution of the orbital semi-major axis:
\begin{equation}
\frac{\dot{a}}{a} = 2
\left[
\frac{\dot{e}}{e}\frac{e^2}{1-e^2}-\frac{I_s\dot{\Omega}_s}{L_{\rm orb}}-\frac{I_p\dot{\Omega}_p}{L_{\rm orb}}
-\frac{I_s\kappa\Omega_s{\rm Min}(\Omega_s,\tilde{\Omega})^2}{L_{\rm orb}}
\right].
\label{eq:dlnadt}
\end{equation}
By integrating this equation we obtain an estimate of the time remaining to spiral-in
for a planet orbiting a slowly-rotating star,
\begin{equation}
t_{\rm remain}=\frac{2 Q'_s}{117n}\frac{M_s}{M_p}\left(\frac{a}{R_s}\right)^5,
\label{eq:tremain}
\end{equation}
giving a result almost identical to the estimate of the same quantity derived from a slightly different formulation by \citet{levrard2008}.
The quantity $a/R_s$ is derived directly from the transit duration for a transiting planet, enabling $t_{\rm remain}$ to be estimated from directly-observed quantities for a given value of $Q'_s$.

\citet{leconte2010} recently suggested that parametrizing the tidal evolution equations in this manner, with the stellar and planetary tidal quality factors constant in time, is not equivalent to the more traditional approach of considering either a constant phase lag or a constant time lag of the tidal bulge. They further suggested that such equations, if truncated to O($e^2$), produce both qualitatively and quantitatively incorrect evolutionary histories for systems with $e>0.2$, and promote the use of equations derived from \citet{hut1981}. Despite using a constant tidal quality factor model the tidal equations that we use in this work are not truncated in this fashion, and thus do not suffer from the problems that \citeauthor{leconte2010} describe.

\section{Computational method}
\label{sec:compute}
An estimate of the value of the scaling constant for magnetic braking, $\kappa$, is required to implement (\ref{eq:ldotwind}). Following the standard Weber-Davis model, the rate of change of the rotation rate of an isolated star owing to magnetic braking can be described by
\begin{equation}
\dot{\Omega}_s = -\kappa\Omega_s^3.
\label{eq:omegadot}
\end{equation}
Integrating this under the assumption that $\Omega_2<<\Omega_1$ and $t_2>>t_1$ gives
\begin{equation}
\kappa =  \frac{\Omega_{2}^{-2}}{2 t_2},
\label{eq:kappa}
\end{equation}which leads to the standard power law for magnetic braking,
\begin{equation}
P \propto t^{\alpha}, 
\label{eq:Sku}
\end{equation} with $\alpha=0.5$ \citep{skumanich1972}. This neglect of the initial conditions is justified by the observed strong convergence of spin rates to a narrow period-colour relation in the Hyades \citep{radick1987}, and in other clusters of similar ages. We set $\Omega_2=\Omega_{s,Hyades}$, calculated by scaling the $P_{rot}$-(J-K) colour relationship found for the Coma-Berenices cluster by \citet{cameron2009} and using J and K magnitudes taken from the SIMBAD on-line data archive, to obtain an estimate for $\kappa$ at the age of the Hyades. This value is assumed to be constant throughout the evolution of the system. For simplicity, in all cases the planet's rotation is initially assumed to be tidally locked such that $\Omega_{p,0}=n$, but $\Omega_p$ is permitted to evolve independently thereafter. 

More recent work has found that the braking law exponent, $\alpha$, diverges slightly from the ideal \citeauthor{skumanich1972} value in some environments (see e.g. \citet{cameron2009}). If we generalise equation\,\ref{eq:omegadot} to raise $\Omega$ to the power $\beta$, then we derive
\begin{equation}
\kappa = \frac{\Omega^{-\frac{1}{\alpha}}}{t/\alpha},
\label{eq:truekappa}
\end{equation}a universal expression for $\kappa$ with $\beta=\frac{1}{\alpha}+1$. We calibrated our implementation of the magnetic braking power law using the stars listed in table\,$4$ of \citet{cameron2009}, finding that an exponent of $\alpha=0.495\pm0.002$ gives good agreement between the measured rotation periods and the evolved stellar rotation rates calculated using our method. We found that there was little to differentiate the evolutionary tracks produced using values of $\alpha$ within this range, and therefore adopted the central value of $\alpha=0.495$.

Starting from a given set of initial conditions $(t_0, e_0, \Omega_{s,0}, \Omega_{p,0}, a_0)$, we integrate the four equations (\ref{eq:dlnedt}), (\ref{eq:dompdt}), (\ref{eq:domsdt}), and (\ref{eq:dlnadt}) using a fourth-order Runge-Kutta scheme, adapted from algorithms in \citet{press1992}. The unit of time is 1\,Gyr, and the integration is allowed to run for the approximate main-sequence lifetime, $t_{\rm MS}\simeq 7(M_s/M_{\odot})^{-3}$\,Gyr for a star of mass $M_s$, or until the planet reaches the Roche limit, as defined by \citet{eggleton1983}. To calculate the stellar radius of gyration the metallicity and mass of the star are used to select an appropriate table from \citet{claret2004,claret2005,claret2006,claret2007}; at each timestep we interpolate through this data using the current time to derive a value for $\alpha_s$. This is not an entirely accurate method, as the series of papers by \citeauthor{claret2004} provides only tables for discrete values of metallicity and stellar mass that in many cases do not coincide with the metallicity of the systems being studied. In these cases we therefore use the table that most closely corresponds to our data.

We use the observed values of $e$, $\Omega_s$ and $a$, and their uncertainties, to evaluate the goodness-of-fit statistic at each timestep according to 
\begin{equation}
\chi^2(t) = \frac{(e_{\rm obs}-e(t))^2}{\sigma^2_e}+\frac{(\Omega_{\rm s,obs}-\Omega_{\rm s}(t))^2}{\sigma^2_{\Omega_s}}+\frac{(a_{\rm obs}-a(t))^2}{\sigma^2_a}.
\end{equation}
To this we add a Bayesian prior on the stellar age $t$ to obtain the statistic 
\begin{equation}
C=\chi^2(t) +\frac{(t_{\rm obs}-t)^2}{\sigma^2_t},
\label{eq:C}
\end{equation}
where $t_{\rm obs}$ and  $\sigma_t$ are the stellar age estimated from the star's density and effective temperature \citep{sozzetti2007}, and its associated uncertainty. The age of the system is taken to be the time of the step at which C is a minimum. 

This forward integration method was built into two separate computational schemes: a grid search that carried out the integration for each node in a four-dimensional grid in $a_0$-$e_0$-$\log(Q'_s)$-$\log(Q'_p)$ parameter space, and a Markov-chain Monte-Carlo (MCMC) optimisation scheme to determine the posterior probability distributions of $\log(Q'_s)$, $\log(Q'_p)$, $e_0$ and $a_0$, and to refine the estimates of the initial parameters returned by the grid search.

The best-fitting evolutionary history investigated by the grid search is taken to be the solution that returns the absolute minimum value of the C statistic, whilst under the MCMC scheme the best-fitting set of parameters are taken to be the median values of the respective posterior probability distributions, with $1-\sigma$ errors derived from the values that delineate the central $68.3$\,percent of the distribution. We choose this approach over the absolute minimum, as the latter strongly depends on the precise sampling of the parameter space by any given Markov chain.

It was assumed that the orbit of the planet would monotonically shrink throughout the integration, thus placing a lower limit on the initial semi-major axis of $a_{0,min}=a_{obs}$. The upper limit on the same parameter was set at $a_{0,max}=0.1$\,AU, a value which encompasses $95$\,percent of the current distribution for transiting planets. The ranges of the tidal quality factors were set to $4.0\leq \log(Q'_p)\leq10.0$ and $5.0\leq \log(Q'_s)\leq10.0$ such that the commonly accepted ranges of values ($10^5 - 10^6$ for $Q'_s$, $10^7 - 10^8$ for $Q'_p$ \citep{baraffe2010}) were encompassed together with an additional section of parameter space.

The initial eccentricity distribution for the ensemble of exoplanets is the subject of much discussion. A value of $e_{0,max}=0.2$ encompasses $93$\,percent of the present distribution and limits our parameter space to the region of validity for our tidal equations, as defined by \citet{leconte2010}. However the migration method by which hot Jupiters find themselves at such short orbital periods is still uncertain. One particular mechanism, Kozai scattering \citep{kozai1962}, can pump the eccentricity to large values, and it has been suggested that short period planets are captured at periastron from such highly eccentric orbits. If this is the case then a much higher upper limit on $e_0$ could be justified.

We must thus take into account the initial times at which we will be starting our integrations. \citet{meibom2005} carried out a survey of tidal circularisation in binary stars. They found that open clusters show a characteristic orbital period at which binaries with the most common initial eccentricity circularise (they define a circularised orbit as one with $e=0.01$). This period was found to vary with the age of the cluster. However it is questionable whether exoplanets, owing to the much lower secondary mass, will tend to circularise at the same orbital period. In fact \citet{hansen2010} applied the formalism of \citet{meibom2005} to exoplanetary systems, finding that circularisation periods were generally much shorter than for stellar binaries. We therefore set $e_{0,max}=0.8$, a value encompassing 99\,percent of the current distribution for transiting exoplanets, to allow for the possibility of eccentric orbits produced by Kozai oscillations.

It is important to note that stellar and planetary obliquities greater than $0.0$ are not considered by the integration methods used herein owing to the algebraic formulation that underlies our computational methods. We therefore assume that the orbital and spin angular momenta vectors are aligned, or anti-aligned. 

\subsection{Markov-chain Monte-Carlo simulation}
\label{sec:MCMC}
Under our MCMC optimisation scheme, the coordinates of the origin for the Markov chain are set to the centre of the parameter space. An exception is made if this lies within $3-\sigma_C$ of the coordinates of the best-fitting initial conditions, in which case the origin coordinates are set to lie outwith the $3-\sigma_C$ contour in parameter space. The burn-in phase was judged to be complete when the minimum value of the test statistic from the current integration exceeded the median of all previous minimum test statistic values for the first time \citep{knutson2008}.

\section{The WASP-18 system}
\label{sec:wasp18}
The hot Jupiter WASP-18\,b orbits the star $\mbox{HD}10069$, an F6 star with an effective temperature $T_{eff}=6400\pm100$\,K and a V magnitude of $9.3$, situated approximately $100$\,pc from the Earth. A fit to stellar isochrones using the model of \citet{girardi2000} gives a stellar age of $0.630^{+0.950}_{-0.530}$\,Gyr, which can be further constrained using the observed lithium abundance. \citet{hellier2009} (hereafter \defcitealias{hellier2009}{H09}\citetalias{hellier2009}) therefore assign the system an age of $0.5-1.5$\,Gyr, a value that the current work attempts to improve upon. WASP-18\,b was the first planet \emph{confirmed} to have a period of less than a day, orbiting its host in just $0.94$\,days. Recent measurements of the Rossiter-McLaughlin effect by \citet{triaud2010} have shown that the system is well aligned; the assumption of spin-orbit alignment required for our analysis is therefore justified. In the absence of a measured rotation period for WASP-18, we assume that the inclination of the stellar axis is $90^o$ such that $\sin\,I=1.0$ and $v_{rot}=v\sin\,I$.

\begin{table}
	\caption{System parameters for WASP-18, taken from \citetalias{hellier2009} with the exception of the rotation period, which is derived from the observed $v\sin\,I$ under the assumption that the inclination of the stellar axis is $90^o$, the J-K colour, which was calculated using data taken from the SIMBAD on-line database, and $\lambda$ \citep{triaud2010}.}
	\label{tab:1}
	\begin{tabular}{lcc}
		\hline \\
		Parameter & Value & Units \\
 		\hline \\
		$M_s$ & $1.25\pm0.13$ & $\mbox{M}_{\mbox{sun}}$ \\[2pt]
		$R_s$ & $1.216^{+0.067}_{-0.054}$ & $\mbox{R}_{\mbox{sun}}$ \\[2pt]
		$J-K$ & $0.278\pm0.032$ & \\[2pt]
		$T_{eff}$ & $6400\pm100$ & K \\[2pt]
		$v\sin\,I$ & $10.77\pm0.04$ & km\,s$^{-1}$ \\[2pt]
		$P_{rot,s}$ & $5.64\pm0.28$ & days \\[2pt]
		System age & $0.630^{+0.950}_{-0.530}$ & Gyr \\[2pt]
		$M_p$ & $10.30\pm0.69$ & $\mbox{M}_{\mbox{jup}}$ \\[2pt]
		$R_p$ & $1.106^{+0.072}_{-0.054}$ & $\mbox{R}_{\mbox{jup}}$ \\[2pt]
		$a$ & $0.02026\pm0.00068$ & AU \\[2pt]
		$e$ & $0.0092\pm0.0028$ & \\[2pt]
		$i$ & $86.0\pm2.5$ & $^o$ \\[2pt]
		$P_{orb}$ & $0.94145299\pm0.00000087$ & days \\[2pt]
		$\lambda$ & $5.0^{+3.1}_{-2.8}$ & $^o$ \\[2pt]
		\hline \\
	\end{tabular}
\end{table}
An observed system age of $0.630\pm0.530$\,Gyr was used for the calculation of the C-statistic for this system, in line with the age found from stellar isochrones by \citetalias{hellier2009}. WASP-18\,b has a radius of the order of that of Jupiter, but is significantly more massive. Lacking a means of calculating $\alpha_p$ however, we set $\alpha_p=\alpha_{Jupiter}$ as a reasonable estimate.

In light of the observed system age being consistent with the age of the Hyades, we compared $P_{rot,s}$ from Table\,\ref{tab:1} to $P_{rot,s}(t=t_{Hyades})$, as calculated for our derivation of $\kappa$ for the system, finding the observed stellar rotation period to be shorter than the expected rotation period of $7.00$\,days at the age of the Hyades. We therefore set $t_0=150$\,Gyr$\approx t_{M35}$, calculating $\Omega_{s,0}$ by scaling the period colour relation of \citet{cameron2009} to this age.

\begin{figure*}
	\includegraphics[width=\textwidth]{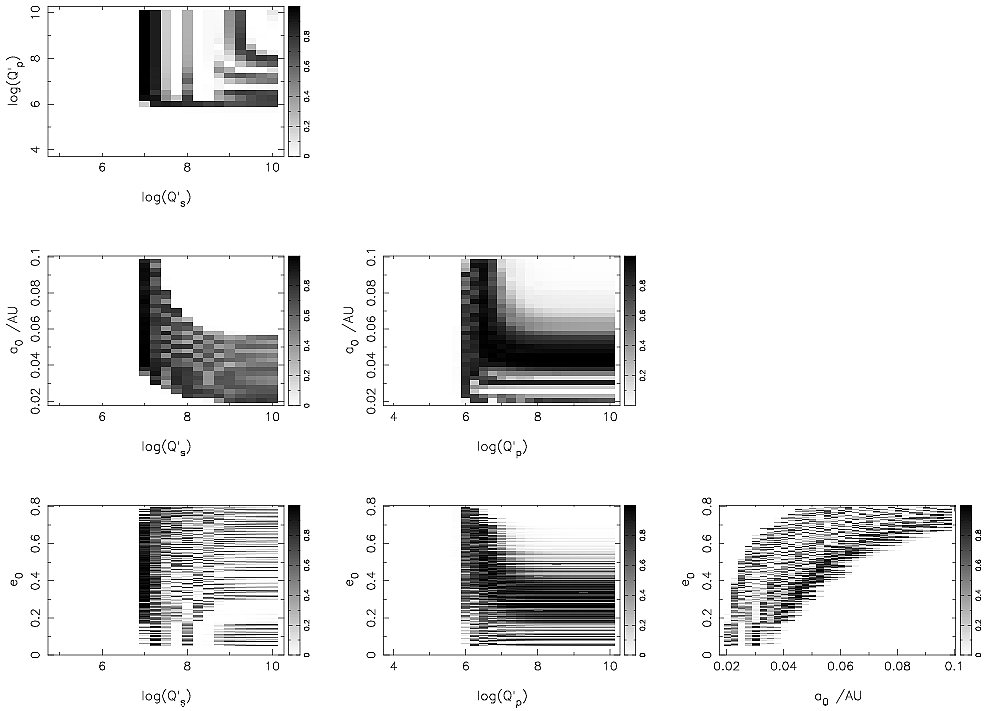}
	\caption{Projection maps of the test statistic probability density for WASP-18. The projections indicate distinct ranges of $\log(Q'_s)$, $\log(Q'_p)$ and $a_0$ that give high probability, corresponding to a low value of the C-statistic. $e_0$ provides a less clear cut high probability region, with several individual values giving good results. The best-fitting initial parameters were found to be $a_0=0.0427$\,AU, $e_0=0.245$, $\log(Q'_s)=7.00$ and $\log(Q'_p)=7.50$, giving a system age of $0.654$\,Gyr.}
	\label{fig:1}
\end{figure*}

The C-statistic values returned by the grid search were converted into probability densities according to
\begin{equation}
P(C) = e^{-\frac{C^2}{2}},
\label{eq:probC}
\end{equation} and plotted as a set of projection maps (Fig.\,\ref{fig:1}). These indicate regions of high probability, and therefore low points in the C-statistic surface, for each pair of initial parameters, and appear to show that there are a range of values for each parameter that could produce plausible evolutionary histories with a good quality of fit. This is particularly noticeable in $e_0$ and $\log(Q'_p)'$, indicating that these parameters have a lesser impact on the orbital and spin evolution of the system than $a_0$ and $\log(Q'_s)$. This is, to a certain extent, unsurprising; the current orbital eccentricity for WASP-18\,b is more uncertain than the orbital separation, and so will have less influence on the value of the C-statistic. It is also interesting to note the generally sharp transitions in Fig.\,\ref{fig:1} between regions with $P(C)>0.4$ and those with $P(C)\approx0$. This indicates that although the values of the initial parameters are somewhat uncertain, they are generally well constrained to a smaller region of parameter space than we have allowed for.

The best-fitting combination of initial parameters in these projection maps was found to be $a_0=0.0428$\,AU, $e_0=0.245$, $\log(Q'_s)=7.00$ and $\log(Q'_p)=7.50$, indicating a system age of $0.654$\,Gyr in good agreement with the existing estimate derived from stellar model fitting. Using the period colour relation of \citet{cameron2009} we calculate an age of $0.387\pm0.024$\,Gyr for WASP-18 using the derived rotation period, inconsistent with both the age from isochrone analysis and the result of our grid search. From the evolutionary track we derive an extremely short remaining lifetime of 0.006\,Gyr, implying that the planet is currently spiralling in towards its host and is on the verge of reaching the Roche limit. This seems incredibly short, and would mean that we have managed to observe WASP-18\,b in a very short window of opportunity. This result is, however, at oodds with the results that we have obtained from our second integration method, the MCMC algorithm.

\begin{figure*}
	\includegraphics[width=\textwidth]{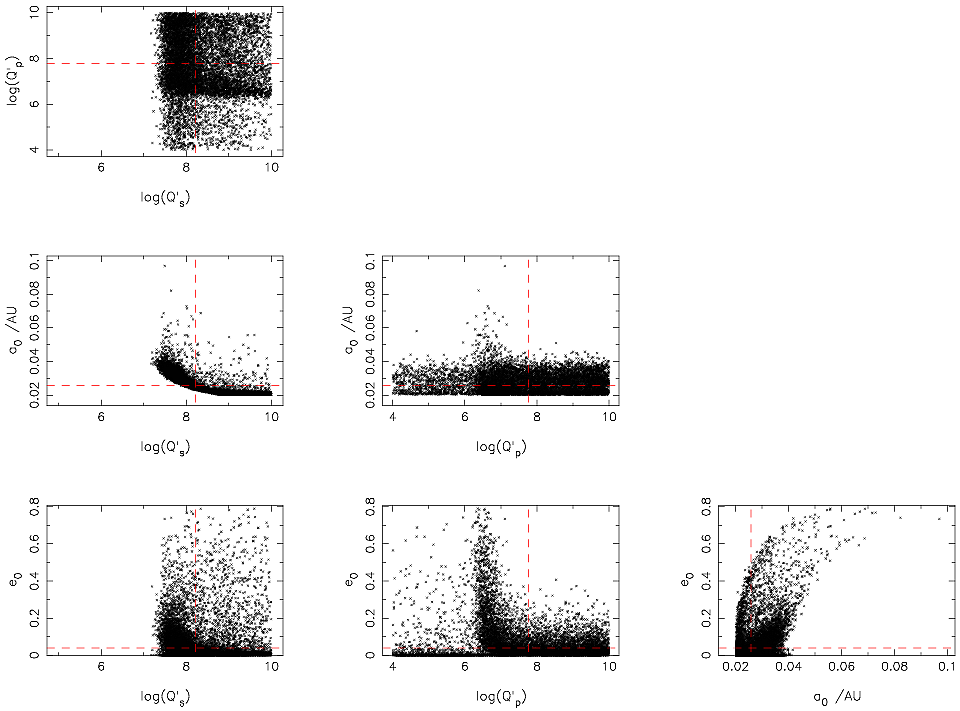}
	\caption{The posterior probability distributions of each pair of jump parameters produced by the MCMC analysis of the WASP-18 system. Note the clear correlations between several of the integration parameter pairs; these arise from the dependence of tidal energy dissipation on the orbital eccentricity, and from the strong coupling that exists between the orbital eccentricity and the orbital separation. The location of the best-fitting parameters in the parameter space explored are denoted by the dashed lines.}
	\label{fig:2}
\end{figure*}

The posterior probability distributions produced by the MCMC integration scheme (Fig.\,\ref{fig:2}) indicate significant correlation between many of the pairs of initial orbital parameters, in particular $a_0$ and $\log(Q'_s)$ although it is also noticeable in several of the other distributions. Some correlation is to be expected; the orbital circularisation time-scale depends on energy dissipation within both the star and the planet \citep{miraldaescude2002}, and so a correlation between the eccentricity and the quality factors that govern tidal energy dissipation is unsurprising. The correlation between $a_0$ and $e_0$ may arise as a result of the strong coupling between eccentricity and separation noted by \citet{jackson2008}. The rapidity with which the orbital separation can decrease is intrinsically linked to the efficiency with which energy is dissipated within the system, a process that is governed by the tidal quality factors. Moreover inspection of (\ref{eq:dlnadt}) indicates that $\dot{a}\propto Q^{-1}$, so the form of the correlation between $a_0$ and $\log(Q'_s)$ that is observed is also as expected. In fact it is surprising that little to no correlation is observed between $a_0$ and $\log(Q'_p)$

\begin{figure}
	\includegraphics[width=0.48\textwidth]{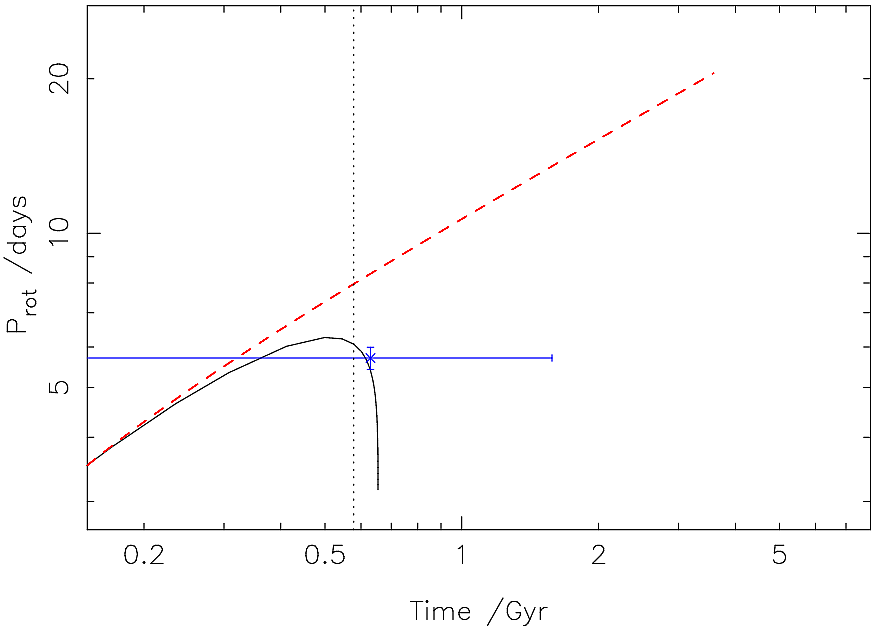}
	\caption{The evolution of the rotation period of WASP-18 from the best-fitting initial conditions found by the MCMC exploration of the allowed parameter space (solid line), along with the evolution that would occur in the absence of tidal interactions (dashed line). For reference, the observed parameters are plotted with associated errors, and the best-fitting age given by the MCMC integration scheme is denoted (dotted line). The tide-governed evolution diverges from the ideal rotational evolution early on, with the effect of tidal interactions on the the rotation of the host star increasing with time until the rotation period dramatically decreases during the final spiral-in of the planetary body.}
	\label{fig:3}
\end{figure}

The best-fitting parameters found using the MCMC code are set out in Table\,\ref{tab:2}, and imply a stellar age of $0.579^{+0.305}_{-250}$\,Gyr, in agreement with the existing age obtained through isochrone fitting, and the age derived from our grid search results. Fig.\,\ref{fig:3} displays the evolution of the stellar rotation for the set of best-fitting parameters; it is important to recall that the age estimate is not evaluated solely on the basis of this parameter, and that the observed orbital eccentricity and semi-major axis play a part as well.

The value of $\log(Q'_s)$ returned by the MCMC-derived integration scheme does not agree with the best-fitting value obtained from the grid search of the defined parameter space. Nor do the values of $a_0$ and $e_0$. We attribute this to the adoption of the set of median parameters from the MCMC scheme to avoid chance encounters with local minima, as well as the discrete nature of the grid search. Comparison of the solution with the lowest value of the C test statistic shows a value of $\log(Q'_s)$ that fits more closely to the grid search result, albeit still in disagreement, although we note that the minimum test statistic value for the grid search is an order of magnitude lower than that for the MCMC exploration of the available parameter space. We also note that although Figs.\,\ref{fig:1} and \ref{fig:2} are broadly similar in form, they differ somewhat in detail. The same correlations between parameters are visible in both figures, but the cutoff in $\log(Q'_s)$ occurs at a slightly greater value in Fig.\ref{fig:2} than Fig.\,\ref{fig:1}. Additionally, the MCMC algorithm explores a more narrow range of both $a_0$ and $e_0$ parameter space than the grid search, but does explore the lower end of $\log(Q'_p)$ space, a region which the grid search suggests gives poor results. We again attribute this to the differences in the manner in which the two algorithms explore the parameter space that we have delineated.

\begin{table}
	\caption{The initial orbital parameters and tidal quality factors of the best-fitting tidal evolution histories produced by the grid search and MCMC integration schemes for the WASP-18 system. The $1-\sigma$ error bars for the MCMC derived values are estimated from the parameter values that encompass the central $68.3$\,percent of the final parameter distributions, and in some cases are inflated by the presence of short `tails' in the distributions.}
	\label{tab:2}
	\begin{tabular}{lcccc}
		\hline \\
		Parameter & Grid search & MCMC & Units \\
 		\hline \\
		$a_0$ & $0.0427$ & $0.0258^{+0.0052}_{-0.0044}$ & AU \\[2pt]
		$e_0$ & $0.245$ & $0.0399^{+0.1023}_{-0.0351}$ & \\[2pt]
		log($Q'_s$) & $7.00$ & $8.21^{+0.90}_{-0.52}$ & \\[2pt]
		log($Q'_p$) & $7.50$ & $7.77^{+1.54}_{-1.25}$ & \\[2pt]
		age & $0.654$ & $0.579^{+0.305}_{-0.250}$ & Gyr \\[2pt]
		$\mbox{t}_{\mbox{remain}}$ & $0.006$ & $0.076^{+0.790}_{-0.044}$ & Gyr \\[2pt]
		\hline \\
	\end{tabular}
\end{table}

Using (\ref{eq:tremain}) with the data from Table\,\ref{tab:2} and the best-fitting value of $\log(Q'_s)$ from the MCMC scheme, we estimate that WASP-18\,b will reach its Roche limit $0.086$\,Gyr from now. The evolutionary track produced using the MCMC results implies a remaining lifetime of $0.076^{+0.742}_{-0.044}$\,Gyr for the planet, consistent with this value. These are, respectively, $2.40$ and $2.12^{+20.73}_{-1.23}$\,percent of the expected main sequence lifetime of the host star, which we estimate to be $3.58$\,Gyr. These remaining lifetimes strongly imply that the WASP-18 planetary system has an short life expectancy, as found by \citetalias{hellier2009}. Although still quite short, these values are far more reasonable than the value of 6\,Myr derived from the grid search.

The evolution of the stellar rotation period (Fig.\,\ref{fig:3}) for the set of initial parameters returned by the MCMC integration scheme indicates that the rotational evolution of the star under the influence of tides initially differs little from our simulation in which the rotation period is governed purely by a Skumanich-type magnetic braking law. However the evolution of the rotation period rapidly diverges from this ideal case, with the rate of spin-down slowing gradually as the star ages and the planet migrates towards its host under the influence of tidal interactions. The final spiral-in of the planetary companion causes a rapid and substantial spin-up of the star; during this process the stellar rotation period is reduced by almost a factor of two from its maximum $6.26$\,day period to $3.16$\,days at the Roche limit, suggesting significant spin up of the host star. Furthermore, Fig.\,\ref{fig:3} clearly shows that, for the age estimate returned by our MCMC integration scheme, the observed rotation period of the host star is quite inconsistent with the rotation period of $7.96$\,days that would be expected at the same age if tidal interactions played no part in the evolution of stellar rotation.

Although the best-fitting parameters that we have adopted show WASP-18\,b in the process of its final spiral in, the true picture is slightly more ambiguous, as evidenced by the large upper uncertainty on the remaining lifetime. Within the $1-\sigma$ parameter ranges there exist combinations of the initial parameters that produce evolutionary tracks of a different form to that displayed in Fig.\,\ref{fig:3}. These evolutionary tracks follow the stellar rotation curve produced by a purely magnetic braking scenario for a much longer period of time, and exhibit little spin up of the host star. Furthermore, in semi-major axis space they indicate that the planet stays at approximately the same semi-major axis for the duration of the star's main sequence lifetime. In these cases the age of the system lies towards the lower end of the adopted range, owing to the necessity of conforming to the calculated stellar rotation period, and the planet has a long remaining lifetime. We note however that these solutions form only a minority of the evolutionary tracks consistent with the parameter ranges that we adopt from the MCMC results. More prevalent were tracks of the same form as that displayed in Fig.\,\ref{fig:3}, but with the spin-up taking place much later such that the stellar rotation implies an age consistent with the gyrochronological estimate and magnetic-braking only scenario.

We attempted to further constrain the range of possible tidal quality factors by visually fitting to the observed parameters and their associated $1-\sigma$ uncertainties. Starting from the best-fitting initial orbital separation and eccentricity, we investigated the evolution of the system across the range of values for both $\log(Q'_s)$ and $\log(Q'_p)$. We found that changing the value of $\log(Q'_p)$ makes no difference to the evolution of the orbital separation or rotation period, but has a strong effect on the eccentricity evolution of the planetary orbit. $\log(Q'_s)$, in contrast, strongly affects the evolution of all integration parameters. Moreover we found that changing the initial semi-major axis, within the adopted range, made a significant difference to the evolution of the stellar rotation period, whilst modifying $e_0$ merely affected the evolution of the orbital eccentricity. We therefore conclude that the two most important parameters with respect to the stellar rotation are $Q'_s$ and $a_0$.

\begin{figure}
	\subfloat{
		\includegraphics[width=0.48\textwidth]{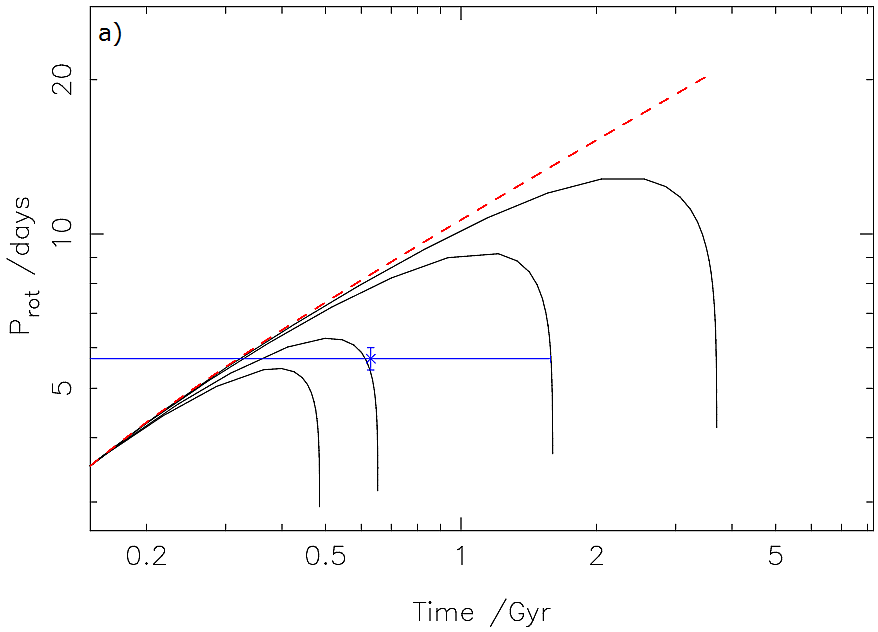}
		\label{fig:4a}}\\
	\subfloat{
		\includegraphics[width=0.48\textwidth]{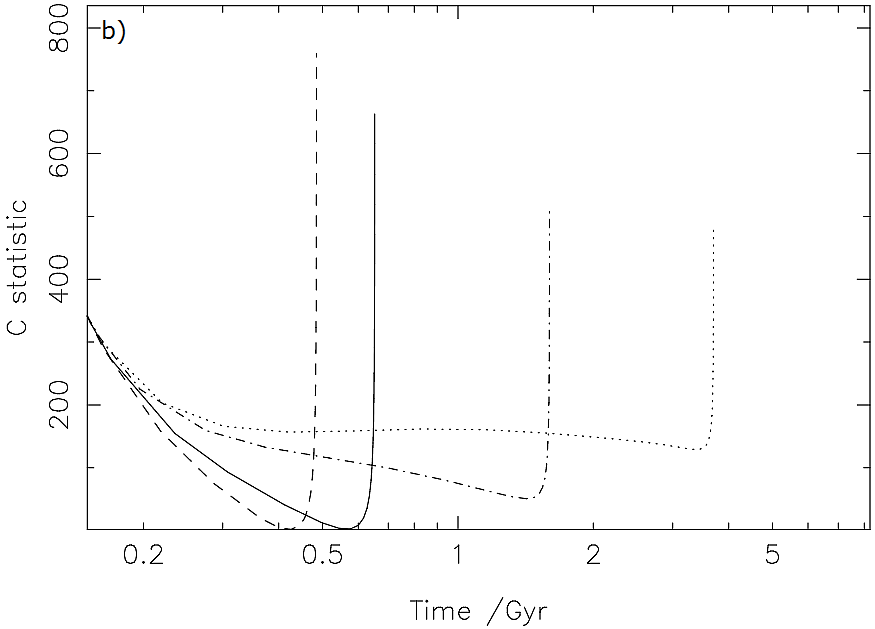}
		\label{fig:4b}}
	\caption{Evolutionary tracks for WASP-18 produced using several different values of $\log(Q'_s)$ within the range consistent with the observed system parameters. The tracks were calculated using the values of $e_0$, $a_0$ and $\log(Q'_p)$ adopted from the MCMC solution and, from left to right (at the end of the track): $\log(Q'_s)=8.02$, $\log(Q'_s)=8.21$, $\log(Q'_s)=8.70$ and $\log(Q'_s)=9.11$. \emph{Upper panel:} The evolution of stellar rotation period with time for WASP-18. The dashed line shows the evolution expected from a Skumanich-type magnetic braking law. Changing the value of $Q'_p$ has no effect on the evolution of the rotation period with time. \emph{Lower panel:} The evolution of the C test statistic with time for the tracks in the upper panel. The time at which C is a minimum is taken to be the age of the system given the set of initial parameters used for that evolutionary track.}
	\label{fig:QrangeW18}
\end{figure}

We were unable to constrain the permissible ranges of the two tidal quality factors any further, owing to the range of values that $a_0$ was able take, but found that for the majority of the possible combinations the rotational evolution of WASP-18\,A gradually diverges from that expected of a purely magnetic braking scenario, and that the star is eventually spun-up by a substantial amount during the final in-spiral of its planetary companion. Fig.\,\ref{fig:4a} shows the rotational period evolution resulting from several values of $\log(Q'_s)$ within the permissible range, with $\log(Q'_p)$, $a_0$ and $e_0$ set to values adopted from the MCMC search.

\section{The WASP-19 system}
\label{sec:wasp19}
The transiting hot Jupiter WASP-19\,b orbits a late G-dwarf star with an orbital period of $0.7888399\pm0.0000008$\,days. The host star has been measured to have $T_{eff}=5500\pm100$\,K, periodic sinusoidal flux variations that indicate a detectable level of intrinsic variability and activity, and a metallicity slightly greater than solar at $\mbox{[M/H]}=0.1\pm0.1$\,dex \citep{hebb2010} (hereafter \defcitealias{hebb2010}{H10}\citetalias{hebb2010}). Measurement of the Rossiter-McLaughlin angle for the system indicates that it is aligned \citep{hellier2011}. 

The age of the system is still somewhat uncertain, with \citetalias{hebb2010} able to determine a constraint of $>1.0$\,Gyr. We note that several studies (e.g. \citet{hansen2010,weidner2010}), as well as the exoplanet encyclopedia, cite an age of $0.6\pm0.5$\,Gyr for the star and attribute it to \citetalias{hebb2010}. We believe that this originates from the abstract of \citetalias{hebb2010} which quotes that age as one of two possibilities. However the text of that paper favours the older age constraint.  For our analysis we used an age of $t_{age}=5.5\pm4.5$\,Gyr, consistent with the isochronal fit and lower bound on the age quoted by \citetalias{hebb2010}, and set $\alpha_p=\alpha_{Jupiter}=0.26401$. Comparing the stellar rotation period from Table\,\ref{tab:3} to the rotation period expected at the age of the Hyades, we found that $P_{rot,s}>P_{rot,Hyades}=8.60$\,days and thus set $t_0=t_{Hyades}$ and $\Omega_{s,0}=\Omega_{s,Hyades}$. 

\begin{table}
	\caption{System parameters and $1-\sigma$ limits for WASP-19, taken from the free eccentricity fit of \citetalias{hebb2010} with the exceptions of the J-K colour, which is derived from data taken from SIMBAD, and the rotation period (Collier Cameron, \emph{priv. comm.}).}
	\label{tab:3}
	\begin{tabular}{lcc}
		\hline \\
		Parameter & Value & Units \\
		\hline\\
		$M_s$ & $0.95\pm0.10$ & $\mbox{M}_{\mbox{sun}}$\\[2pt]
		$R_s$ & $0.93^{+0.05}_{-0.04}$ & $\mbox{R}_{\mbox{sun}}$\\[2pt]
		$J-K$ & $0.43$ & \\[2pt]
		$T_{eff}$ & $5500\pm100$ & K \\[2pt]
		$v\sin\,I$ & $4.0\pm0.2$ & km\,s$^{-1}$ \\[2pt]
		$P_{rot}$ & $10.5\pm0.2$ & days \\[2pt]
		System age & $5.5^{+9.0}_{-4.5}$ & Gyr \\[2pt]
		$M_p$ & $1.14\pm0.07$ & $\mbox{M}_{\mbox{jup}}$\\[2pt]
		$R_p$ & $1.28\pm0.07$ & $\mbox{R}_{\mbox{jup}}$\\[2pt]
		$a$ & $0.0164^{+0.0005}_{-0.0006}$ & AU\\[2pt]
		$e$ & $0.02^{+0.02}_{-0.01}$ & \\[2pt]
		$i$ & $80.8\pm0.8$ & \\[2pt]
		$P_{orb}$ & $0.7888399\pm0.0000008$ & days \\[2pt]
		\hline\\
	\end{tabular}
\end{table}

\begin{figure*}
	\includegraphics[width=\textwidth]{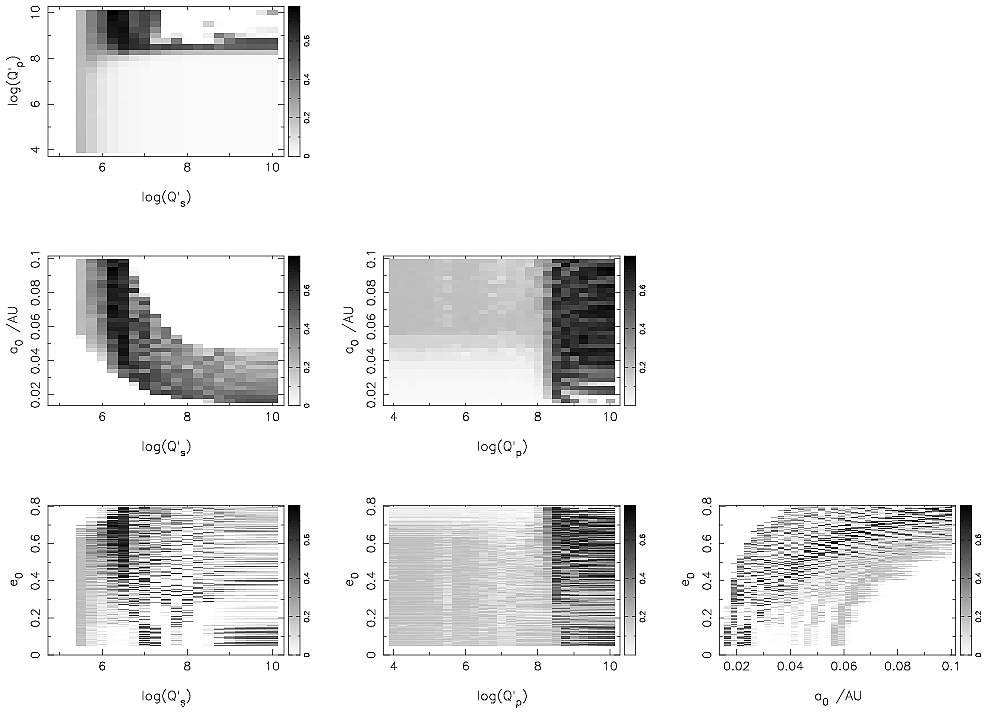}
	\caption{C-statistic probability density projection maps for WASP-19\,b. A small region of high probability is clearly visible in $\log(Q'_s)$ and $\log(Q'_p)$ parameter space, but in $a_0$ and $e_0$ the range of values that give high probability solutions are much broader. The point of maximum probability, indicating the minimum in the C-statistic surface, was found to occur at parameter space coordinates of $a_0=0.0939$\,AU, $e_0=0.735$, $\log(Q'_s)=6.25$ and $\log(Q'_p)=10.0$, giving a system age of $2.11$\,Gyr.}
	\label{fig:5}
\end{figure*}

Fig.\,\ref{fig:5} shows the C-statistic probability density projection maps produced from the grid search results for the WASP-19 system. They indicate a fairly small region of high probability for $\log(Q'_s)$ and $\log(Q'_p)$, but appear to show that there are broader ranges of both $e_0$ and $a_0$ that give strong solutions. The absolute maximum probability, and thus the minimum value of the C statistic, was found to occur when the initial conditions were $a_0=0.0939$\,AU, $e_0=0.735$, $\log(Q'_s)=6.25$ and $\log(Q'_p)=10.0$, giving a system age of $2.11$\,Gyr that is consistent with the existing constraint of $>1.0$\,Gyr from \citetalias{hebb2010}. Using the period-colour relation of \citet{cameron2009} we calculate a gyrochronological system age of $0.899\pm0.037$\,Gyr, immediately suggesting that gyrochronological analysis is inappropriate for this system.

Comparing the sets of projection maps for the two systems, it is apparent that there are similarities in the forms of Figs.\,\ref{fig:1} and \ref{fig:5}, particularly in the $a_0-\log(Q'_s)$ and $e_0-a_0$ maps. However the range of probability values covered by the greyscale is slightly narrower for WASP-19, where the maximum probability is approximately $0.8$, compared to WASP-18, where the maximum probability is almost $1.0$. From this we deduce that the best-fitting orbital solution for WASP-19 is less certain than our best-fitting grid search solution for WASP-18. This lower maximum probability, coupled with the form of the projection maps, also suggests that the range of tidal quality factors for which valid orbital solutions exist is greater for this system than it was for WASP-18.

\begin{figure*}
	\includegraphics[width=\textwidth]{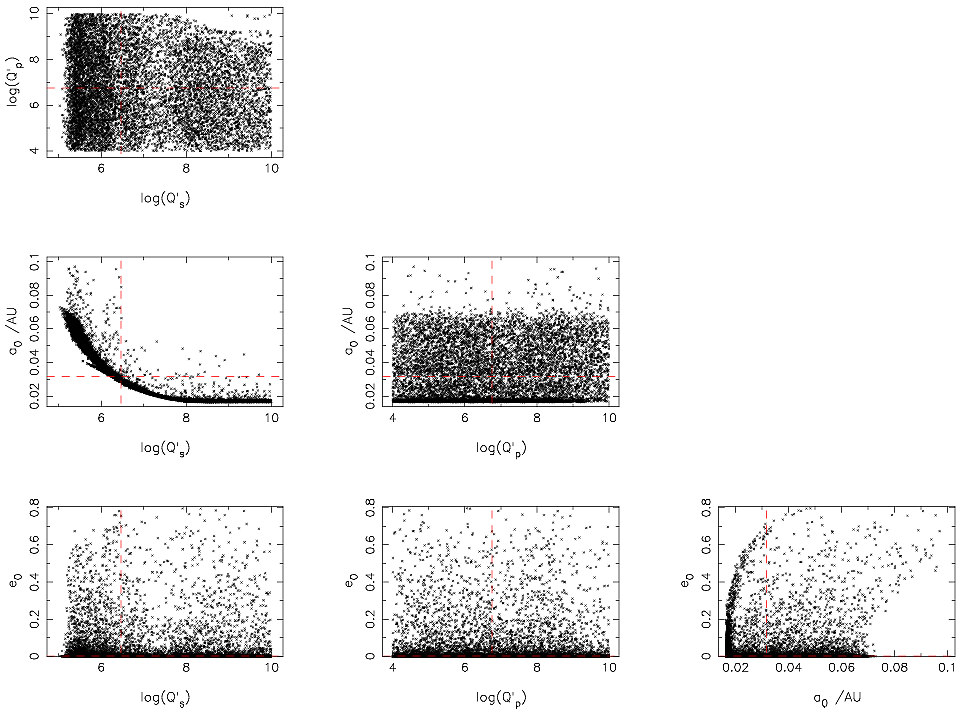}
	\caption{The posterior probability distributions of each pair of integration parameters produced by the MCMC analysis of the WASP-19 system. The correlations between the parameters are substantially different than those for the WASP-18 system, with that between $\log(Q'_s)$ and $a_0$ being the only particularly noticeable correlation. The location of the best-fitting parameters in the parameter space explored are denoted by the dashed lines.}
	\label{fig:6}
\end{figure*}

The posterior probability distributions produced by the MCMC algorithm exhibit a strong correlation between $a_0$ and $\log(Q'_s)$, of a similar form to that observed for the WASP-18 system albeit extended to lower values of $\log(Q'_s)$. Also present is a correlation between $e_0$ and $a_0$ that is somewhat similar to that apparent in Fig.\,\ref{fig:2}, but the other correlations that were present in the distributions for WASP-18 are absent for WASP-19. The region of parameter space explored by the MCMC algorithm is also substantially greater for the WASP-19 system than it was for the WASP-18 system, although this is not unexpected after the comparative appearance of Figs.\,\ref{fig:1} and \ref{fig:5}.

It is also interesting to note the relative lack of agreement between the results from the grid search and the MCMC search methods for this system; the only search parameter for which the two sets of results agree is $\log(Q'_s)$. This is also apparent from inspection of Figs\,\ref{fig:5} and \ref{fig:6}, as the region of greatest probability in the former does not match the location of the adopted parameters in the latter. As for the WASP-18 system, we attribute this to the adoption of the set of median parameters from the MCMC scheme; comparison of the solution with the lowest value of the C test statistic shows that it more closely fits with the location of the high probability region in Fig.\,\ref{fig:5}. Fortunately the stellar tidal quality factor, the most important parameter when considering the evolution of the stellar rotation, is relatively well constrained in our MCMC-derived solution.

The probability distribution for the initial orbital eccentricity of the WASP-19 system also shows weaker clustering than for the WASP-18 system, and the clustering that is present encompasses a much smaller range of eccentricity values above $e_0=0$. This may be symptomatic of the fact that the eccentricity of the system is not well known; previous analysis of the parameters of the system found little difference in the fit to the observed transit lightcurve between the cases in which the orbit was forced to be circular, and in which the eccentricity was allowed to float \citepalias{hebb2010}. In the latter case, the best-fitting eccentricity value was only $0.02^{+0.02}_{-0.01}$, and it is this value that we used in our $C$ statistic calculations.

\begin{table}
	\caption{The initial orbital parameters and tidal quality factors of the best-fitting tidal evolution histories produced by the grid search and MCMC integration schemes for the WASP-19 system. The $1-\sigma$ error bars for the MCMC derived values are estimated from the parameter values that encompass the central $68.3$\,percent of the final parameter distributions.}
	\label{tab:4}
	\begin{tabular}{lcccc}
		\hline \\
		Parameter & Grid search & MCMC & Updated & Units \\
 		\hline \\
		$a_0$ & $0.0939$ & $0.0317^{+0.0228}_{-0.0146}$ & $0.0317^{+0.0228}_{-0.0089}$ & AU \\[2pt]
		$e_0$ & $0.735$ & $0.0017^{+0.0597}_{-0.0016}$ & & \\[2pt]
		log($Q'_s$) & $6.25$ & $6.47^{+2.19}_{-0.95}$ & $6.47^{+0.67}_{-0.95}$ & \\[2pt]
		log($Q'_p$) & $10.0$ & $6.75^{+1.86}_{-1.77}$ & & \\[2pt]
		age & $2.11$ & $1.60^{+2.84}_{-0.79}$ & & Gyr \\[2pt]
		$\mbox{t}_{\mbox{remain}}$ & $0.0038$ & $0.0067^{+1.1073}_{-0.0061}$ & & Gyr \\[2pt]
		\hline \\
	\end{tabular}
\end{table}

\begin{figure}
	\includegraphics[width=0.5\textwidth]{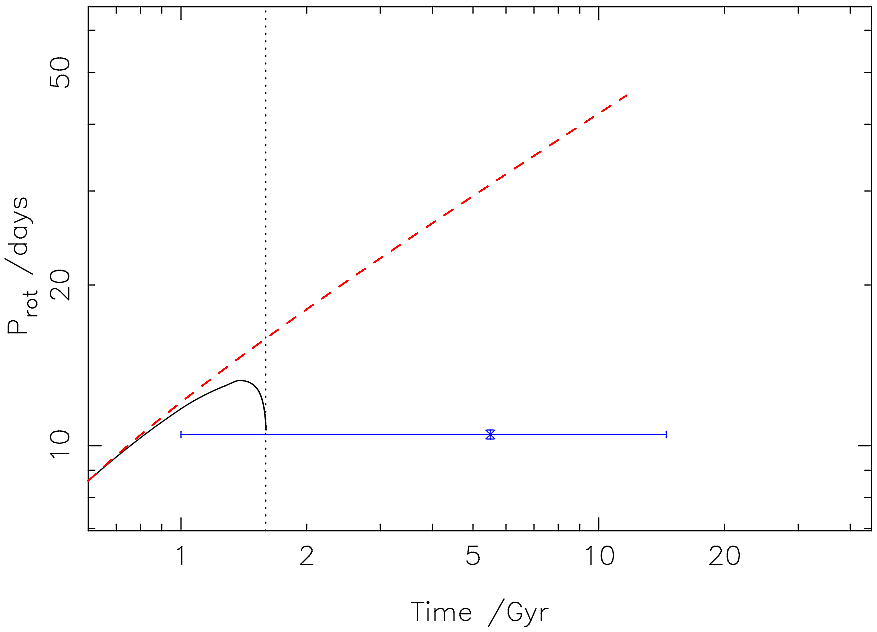}
	\caption{The evolution of the rotation period of WASP-19 from the best-fitting initial conditions found by the MCMC exploration of the allowed parameter space, along with the evolution that would occur in the absence of tidal interactions (dashed line). The tide-governed evolution follows the gradual spin-down of the isolated evolution closely until spiral-in of the planet begins, at which point the stellar rotation period rapidly decreases. The observed rotation period is plotted with associated errors, and implies that WASP-19\,b is in the final spiral-in stage of its orbital evolution. Furthermore it is irreconcilable with the expected spin down from magnetic braking alone, and can only be explained by invoking spin-up during tidal interactions.}
	\label{fig:7}
\end{figure}

The best-fitting parameters given by the MCMC integration scheme, set out in Table\,\ref{tab:4}, imply a stellar age of $1.60^{+2.84}_{-0.79}$\,Gyr, in broad agreement with the loose constraint of $\mbox{age}>1.0$\,Gyr found by \citetalias{hebb2010}. The gyrochronological age that we have calculated for the system also agrees with this age estimate but lies at the lower end of the range, suggesting that gyrochronology provides a possible, if unlikely estimate for the age of WASP-19\,A. The data in Table\,\ref{tab:4} further imply a remaining lifetime for the planet of $0.0067^{+1.1073}_{-0.0061}$\,Gyr, a mere $0.4$\,percent of the estimated system age. This strongly suggests that WASP-19\,b is in the final, spiral-in stage of its orbital evolution, a conclusion supported by the stellar rotation period evolutionary track that results from integration of the best-fitting parameters (Fig.\,\ref{fig:7}). As with the WASP-18 system we attempted to further constrain the range of possible tidal quality factors by visually fitting to the observed parameters; in this case we were able to reduce the upper limit on $\log(Q'_s)$ and raise the lower limit on $a_0$. The updated values are given in Table\,\ref{tab:4}. Updating the limits on these parameters have no effect on the $1-\sigma$ limits of the age or remaining lifetime.

The observed spin rate of WASP-19\,b is inconsistent with evolution governed only by magnetic braking; Fig.\,\ref{fig:7} shows that, at the system age implied by our MCMC solution, the observed period of $10.5\pm0.2$\,days is substantially less than the period of $15.9$\,days implied by our magnetic braking only model. It is possible that WASP-19\,b is in the process of spiraling-in to the Roche limit, spinning up its host as it does so.

\begin{figure}
	\includegraphics[width=0.5\textwidth]{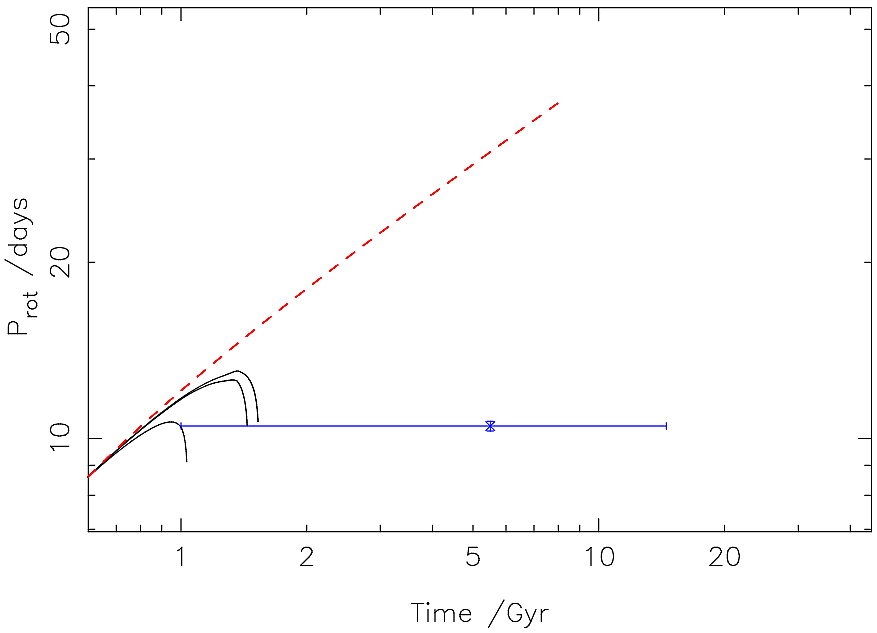}
	\caption{The evolution of stellar rotation period with time for WASP-19 for a range of $\log(Q'_s)$ values. The dashed line shows the evolution expected from a Skumanich-type magnetic braking law, whilst the three solid lines are the evolution that results from the best-fitting $e_0$, $a_0$ and $\log(Q'_p)$ given by the MCMC algorithm, and, from left to right (at the observed system age), $\log(Q'_s)=6.14$, $\log(Q'_s)=6.47$ and $\log(Q'_s)=6.52$. These tracks imply that WASP-19\,b is very close to the Roche limit, and will reach it in less than 10\,Myr.}
	\label{fig:8}
\end{figure}

Fig.\,\ref{fig:8} shows the rotational period evolution resulting from $\log(Q'_s)$ values within the $1-\sigma$ limits returned by the MCMC search scheme. These reinforce the idea that WASP-19\,b is on the verge of reaching its Roche limit; for the best-fitting tidal quality factors the end of the evolutionary track is at $P_{rot}=10.5$\,days compared to the observed $P_{rot}=10.5\pm0.2$\,day, placing the planet precisely on the Roche limit and suggesting that we have been fortunate to observe WASP-19\,b at all.

Much as with the WASP-18 system though, this is not the entire picture presented by our results. The upper limit on $t_{remain}$ implied by our adopted solution is $1.114$ \,Gyr, a significant 69.6\,percent of the estimated age and 13.7 percent of the MS lifetime of the host star. This strongly suggests that there exist solutions that more closely follow the stellar rotation evolution expected when only magnetic braking is acting on the star. To investigate this we characterised the orbital evolutionary tracks for each $\log(Q'_s)-a_0$ node from the grid search that was consistent with our results from the MCMC algorithm. Since our investigation into WASP-18\,b's future evolution led us to the conclusion that $e_0$ and $\log(Q'_p)$ have little effect on the evolution of the stellar rotation, we set these parameters to the MCMC values from table\,\ref{tab:4}.

We found that for each value of $a_0$ there was only a very narrow range of $\log(Q'_s)$ values that gave solutions with an age $\geq1.0$\,Gyr \emph{and} a rotation period at that age consistent with the observed value. In fact only $13$\,percent of the nodes investigated gave such solutions. The majority of the nodes gave solutions that either did not reach an age of $1.0$\,Gyr, instead showing the distinctive spiral-in signature at younger ages, or that followed the magnetic braking-only track for much longer such that the spin-up induced by the final spiral-in was insufficient to reduce the stellar rotation period back to the observed value. Tracks that fell into this latter category often returned ages consistent with our gyrochronological calculation, although some did return older ages consistent with the $1.0$\,Gyr lower limit owing to the influence of the orbital eccentricity and semi-major axis.

This raises questions concerning WASP-19\,A's true age. There is evidence from isochrone fitting that the star is older than 1\,Gyr. This is supported by the investigation of \citetalias{hebb2010} into the space velocity of the star, which found that 65\,percent of simulated stars with similar stellar properties in a small volume around WASP-19\,A were older than 1.0\,Gyr. But the rotation period of WASP-19\,A implies a younger age of $0.899$\,Gyr through gyrochronology. The lithium abundance of $\log(A[Li])<1.0$ quoted by \citetalias{hebb2010}, while consistent with either estimate as it only constrains the age to greater than $0.6$\,Gyr, would appear to support a younger age, as the presence of any lithium tends to rule out a more evolved star. There therefore appear to be two main possibilities; either the star is old and has been spun up by the infall of the planet, or the star is younger, still following its natural spin down and more dense and/or hotter than expected for its age. Both of these are consistent with the results that we have obtained from our search methods, so which is the more plausible?

\subsection{Is W19 young or old?}
\label{sec:W19dicuss}
We consider our results for WASP-19 in the context of a larger ensemble of exoplanetary systems with similar stellar and planetary properties, assuming that the value of $\log(Q'_s)=6.47$ adopted for WASP-19 is applicable to the other systems under consideration. We find that for 7 of the 9 planetary systems in the sample the remaining lifetime is $<1$\,Gyr. Furthermore we find that WASP-4\,b has a remaining lifetime of $0.017$\,Gyr, and calculate that it lies at a distance of only 1.4 times the Roche limit from its host star; this places it in apparently similar circumstances to WASP-19\,b. WASP-2\,b and WASP-10\,b have remaining lifetimes at least one order of magnitude greater than those of the rest of the sample, but they also have significantly longer orbital periods.

\begin{table*}
	\begin{threeparttable}
	  \caption{A comparison of the WASP-19 system to a sample of transiting hot Jupiter systems with $M_p>M_J$ in close orbits around stars similar to, or cooler than, WASP-19\,A. We calculate the remaining lifetime for each planet using equation\,\ref{eq:tremain}, assuming that the stellar tidal quality factor of $\log(Q'_s)=6.47$ adopted for WASP-19\,A is applicable across the entire sample. Note that the remaining lifetime for WASP-19\,b quoted here does not agree with the value derived from the evolutionary track displayed in Fig.\,\ref{fig:7}.}
	  \label{tab:5}
	  \begin{tabular}{lllllllll}
	  	\hline \\
	  	System & $\mbox{M}_p$ /$\mbox{M}_J$ & $\mbox{R}_p$ /$\mbox{R}_J$ & $\mbox{M}_s$ /$\mbox{M}_\odot$ & $\mbox{R}_s$ /$\mbox{R}_\odot$ & $a$ /AU & $P$ /days & $\mbox{t}_{remain}$ /Gyr & Age /Gyr\\
	  	\hline \\
	  	WASP-19 & $1.14$ & $1.28$ & $0.95$ & $0.93$ & $0.0164$ & $0.7888399$ & $0.012$ & $1.60^{+2.84}_{-0.79}$ \\[2pt]
	  	WASP-2 & $0.847$ & $1.079$ & $0.84$ & $0.834$ & $0.03138$ & $2.1522254$ & $1.708$ & $11.9^{+8.1}_{-4.3}$ \tnote{a} \\[2pt]
	  	WASP-4 & $1.12$ & $1.416$ & $0.93$ & $1.365$ & $0.023$ & $1.3382282$ & $0.017$ & $7.0^{+5.2}_{-4.5}$ \tnote{a} \\[2pt]
	  	WASP-5 & $1.63$ & $1.171$ & $1.00$ & $1.15$ & $0.02729$ & $1.6284246$ & $0.080$ & $3.0\pm1.4$ \tnote{b} \\[2pt]
	  	WASP-10 & $3.06$ & $1.08$ & $0.71$ & $0.783$ & $0.0371$ & $3.0927616$ & $1.818$ & $0.8\pm0.2$ \tnote{c} \\[2pt]
	  	CoRoT-1 & $1.03$ & $1.49$ & $0.95$ & $1.11$ & $0.0254$ & $1.5089557$ & $0.093$ & \\[2pt]
	  	CoRoT-2 & $3.31$ & $1.465$ & $0.97$ & $0.902$ & $0.0281$ & $1.7429964$ & $0.159$ & \\[2pt]
	  	OGLE-TR-113 & $1.24$ & $1.11$ & $0.78$ & $0.77$ & $0.0229$ & $1.4324772 $ & $0.224$ & $>0.7$ \tnote{d}\\[2pt]
	  	TrES-3 & $1.91$ & $1.305$ & $0.92$ & $0.813$ & $0.0226$ & $1.30618608$ & $0.111$ & $0.1^{+0.7}_{-0.0}$ \tnote{a} \\[2pt]
	  	\hline \\
	  \end{tabular}
		\begin{tablenotes}
			\item(a)\citet{southworth2010} \item(b)\citet{anderson2008} \item(c)\citet{christian2009} \item(d)\citet{melo2006}
		\end{tablenotes}
	\end{threeparttable}
\end{table*}

It is interesting to consider the range of the stellar tidal quality factor that produces certain threshold values of $\mbox{t}_{remain}$. We find that, if we exclude the longer period WASP-2 and WASP-10 systems, $\log(Q'_s)>7.10$ leads to an estimated $\mbox{t}_{remain}\geq1.0$\,Gyr for all planets in the sample, whilst a remaining lifetime greater than $0.1$\,Gyr requires $\log(Q'_s)>6.12$. For WASPs -2 and -10 we find that the values are $6.24$ and $5.24$, and $6.22$ and $5.22$ respectively.

The tidal evolution solution that we presented previously would seem to support the hypothesis that the star is old, with stellar spin up accounting for the rotation period. But the results in Table\,\ref{tab:5} would seem to add weight to the idea that the system is actually younger than expected. The likelihood of observing one system in such a condition is low, but to observe two such systems as Table\,\ref{tab:5} implies that we have done with WASPs -4 and -19, seems incredulous.

\begin{figure}
	\includegraphics[width=0.5\textwidth]{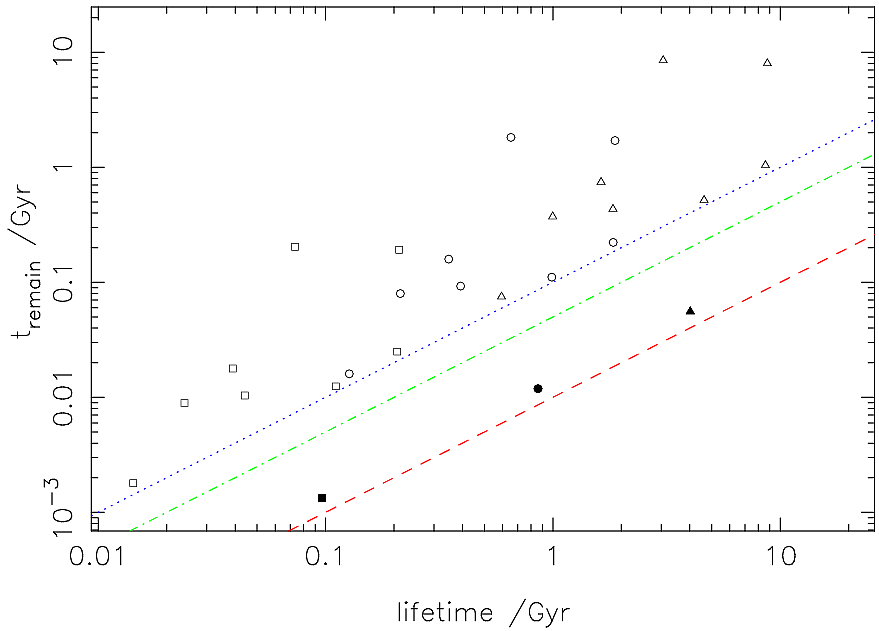}
	\caption{Remaining lifetime as a function of total planetary lifetime for the sample of transiting planets in Table\,\ref{tab:5}, as calculated using \ref{eq:tremain} assuming that the values of $\log(Q'_s)$ and $a_0$ adopted for the WASP-19 system from our MCMC results apply universally. We plot the results for $\log(Q'_s)=5.52$\,(squares), $\log(Q'_s)=6.47$\,(circles) and $\log(Q'_s)=7.14$\,(triangles). The data for WASP-19\,b are denoted by filled symbols. The lines represent $t_{remain}=1$\,percent (red, dash), $5$\,percent (green, dot-dash) and $10$\,percent (dark blue, dot) of the total planetary lifetime. WASP-19\,b is clearly separate from the rest of the sample, indicating that it is in a unique situation.}
	\label{fig:9}
\end{figure}

Considering the remaining lifetimes in the context of the total planetary lifetime suggests a different scenario. \ref{eq:tremain} can also be used to calculate the total planetary lifetime if the initial semi-major axis is known. We calculated the total planetary lifetimes assuming that the value adopted for WASP-19\,b, $a_0=0.0317$\,AU applied unilaterally across our sample, and in Fig.\,\ref{fig:9} plot remaining lifetime as a function of total lifetime. It is immediately apparent that WASP-19\,b is a special case, being clearly separated from the rest of the sample. The remaining lifetime of WASP-19\,b is only between 1 and 2\,percent of the total lifetime, whilst for all of the other planets in the sample this figure is greater than 10\,percent. The short remaining lifetime of WASP-4\,b is thus somewhat misleading, as it represents a significant portion ($\approx12.6$\,percent) of the total lifetime of the planet.

\begin{figure}
	\includegraphics[width=0.5\textwidth]{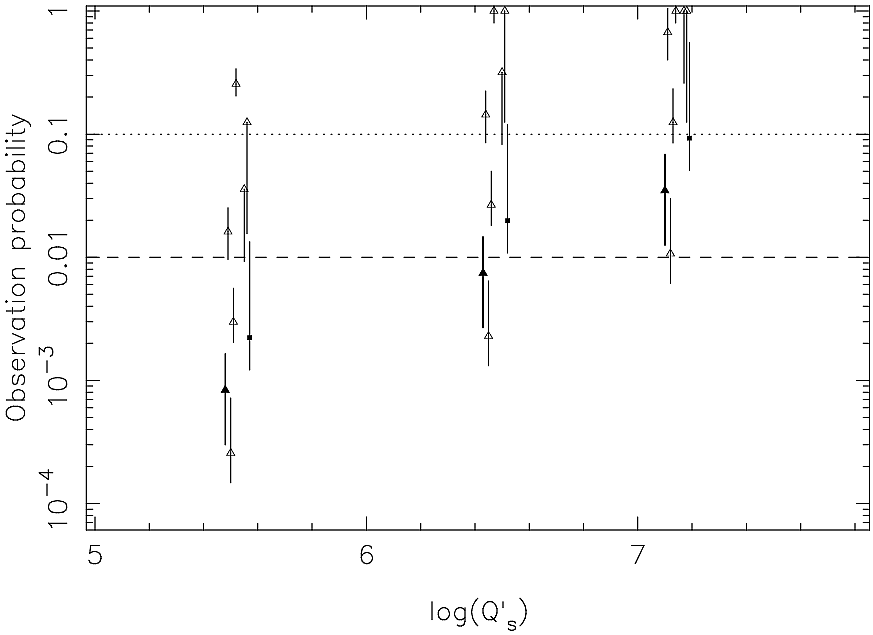}
	\caption{The probability of observing the systems in Table\,\ref{tab:5} in their present configuration, as a function of $\log(Q'_s)$ within the range encompassed by the $1-\sigma$ errors from the MCMC results. Systems without a stellar age in the literature were disregarded. The data are offset slightly from the true value of $\log(Q'_s)$ to aid clarity for the error bars. Systems for which $P>1$ are plotted at $P=1$. WASP-19 is denoted by filled symbols; triangles for the probability using the age estimate from the MCMC results, squares for the probability calculated using the estimate of $0.6\pm0.5$\,Gyr prevalent in the literature. At $\log(Q'_s)=6.47$ two systems, including WASP-19 with its older age, show a probability of observation that is less than 1\,percent. Increasing the stellar tidal quality factor increases the probability that we have observed the systems in the state implied by their remaining lifetimes.}
	\label{fig:10}
\end{figure}

There are several ways in which we can analyse the remaining lifetimes calculated here, one of which is to examine the probability that we have managed to observe the system in its present configuration. We calculated this for the systems in Table\,\ref{tab:5}, disregarding those with no literature age estimate. For WASP-19 we calculate the probability using both our own age adopted age estimate, and the younger age of $0.6\pm0.5$ that can be found in the literature. Fig.\,\ref{fig:10} displays the results for a range of $\log(Q'_s)$ values consistent with the $1-\sigma$ limits derived from the MCMC posterior probability distributions. For the value of $\log(Q'_s)$ adopted from the MCMC solution we find that two systems, including WASP-19 at our older age estimate, have a probability of observation of less than 1\,percent. Using the younger age estimate for WASP-19 pushes the probability up to 2\,percent, which is more plausible (see analysis in \citet{hellier2011}). Increasing the value of $\log(Q'_s)$ to our upper limit of 7.14 increases the probability of observing WASP-19\,b to approximately 3.5\,percent for the older age, and approximately 9\,percent for the younger age. However if we consider the lower limit of $\log(Q'_s)=5.52$ then four systems, including both ages of WASP-19, have observation probabilities less than 1\,percent. This is an entirely implausible situation.

However this analysis assumes that all of the planets in our sample started at the same distance from their respective host stars, and that they experience tidal interactions of the same strength. It seems somewhat unlikely that this accurately represents reality, as the planetary systems in our sample show significant variation in their properties. Previous studies \citep{matsumura2010,hansen2010} have found that different planetary systems are likely to experience different strengths of stellar and planetary tide, so describing these systems with a single value is almost certainly unphysical. This would, of course, mean that the observational probabilities for several of these hot Jupiters could be significantly greater.

To further muddy the waters we add evidence for an old age by carrying out additional stellar model fits. We used the updated Padova models \citep{marigo2008}, which imply an age of $7.94^{+4.65}_{-2.93}$\,Gyr, and the latest version of the Yonsei-Yale isochrones \citep{demarque2004}, which return an age of $7^{+2}_{-3}$\,Gyr. Both estimates agree with the earlier isochrone fit of \citetalias{hebb2010}. We also investigated the values of $T_{eff}$ and $\rho_s$ required for a stellar age of $0.6\pm0.5$\,Gyr in these two models, assuming the same metallicity; the results are given in Table\,\ref{tab:6}. The correction required to the effective temperature if the stellar density is correct is not improbable, but the converse is not necessarily true. Modifying the stellar density requires a change to either the stellar mass or radius, or both, either of which could have a dramatic effect on our estimates of the planetary parameters. This is notwithstanding the more likely case, which is that both the temperature and density are in need of a slight adjustment. A younger age for the star does not seem, therefore, outwith the realms of possibility.

For completeness we also attempted to force an age of 0.6\,Gyr by adjusting the stellar metallicity whilst leaving the effective temperature and stellar density at the values from \citetalias{hebb2010}. We found that the upper limit of $Z=0.03$ imposed on the Padova isochrones prevented us from reaching such a low age, leaving us with a lowest age estimate of $3.98\pm3.96$\,Gyr, which could be consistent with a star the age of the Hyades. When fitting to the Yonsei-Yale isochrones we found that a value of $[Fe/H]=0.32$ gave an age of $2.5^{+3.5}_{-1.9}$\,Gyr, just consistent with the younger age that we were looking for. Fitting to the 0.6\,Gyr isochrone itself would require a much greater metallicity, but given that this value of the iron abundance is already 0.05\,dex greater than the maximum elemental abundance found by \citetalias{hebb2010} (specifically the upper $1-\sigma$ limit on the Calcium abundance), it is clear that such a fit would be utterly inconsistent with the current spectral analysis.

It therefore seems that a combination of slightly increased elemental iron abundance, greater stellar density and higher effective temperature could produce stellar model fits more consistent with the young stellar age suggested by gyrochronology. Indeed if one is using the Padova models then such a combination lies just within the existing $1-\sigma$ ranges for the three parameters.

\begin{table}
	\caption{The age estimates obtained through fitting the stellar parameters of WASP-19\,A to stellar models. Adjusting the stellar density, effective temperature or iron abundance in isolation lowers the age that the model fit returns. Obtaining an age of $0.6$\,Gyr required the parameters to be increased beyond their $1-\sigma$ limits when done in isolation. If adjusted as a set, the required age could be obtained with more plausible values.}
	\label{tab:6}
	\begin{tabular}{lllll}
	\hline \\
	Model & $\rho_s /\rho_\odot$ & $T_{eff}$ /K & [Fe/H] & age /Gyr \\
	\hline \\
	Padova & $1.13\pm0.12$ & $5500\pm100$ & $0.02$ & $7.94^{+4.65}_{-2.93}$ \\[2pt]
	Padova & $1.45\pm0.10$ & $5500\pm100$ & $0.02$ & $0.631^{+3.35}_{-0.627}$ \\[2pt]
	Padova & $1.13\pm0.12$ & $5805\pm100$ & $0.02$ & $0.631^{+3.35}_{-0.626}$ \\[2pt]
	Padova & $1.13\pm0.12$ & $5500\pm100$ & $0.2$ & $3.98\pm3.96$ \\[2pt]
	Padova & $1.25\pm0.10$ & $5585\pm100$ & $0.12$ & $0.631^{+3.35}_{-0.625}$ \\[2pt]
	Yonsei-Yale & $1.13\pm0.12$ & $5500\pm100$ & $0.02$ & $7^{+2}_{-3}$ \\[2pt]
	Yonsei-Yale & $1.52\pm0.10$ & $5500\pm100$ & $0.02$ & $0.60^{+1.90}_{-0.54}$ \\[2pt]
	Yonsei-Yale & $1.13\pm0.12$ & $5885\pm100$ & $0.02$ & $0.60^{+1.90}_{-0.54}$ \\[2pt]
	Yonsei-Yale & $1.13\pm0.12$ & $5500\pm100$ & $0.32$ & $2.5^{+3.5}_{-1.9}$ \\[2pt]
	Yonsei-Yale & $1.30\pm0.10$ & $5625\pm100$ & $0.12$ & $0.60^{+1.90}_{-0.40}$ \\[2pt]
	\hline \\
	\end{tabular}
\end{table}

As noted previously, the constraint placed on the age by the lithium abundance has the potential to rule out the young stellar age. The detection of a substantial lithium abundance in a stellar spectrum strongly implies a young age for the star, even one that hosts planets, as recent work by \citet{baumann2010} shows that there is no correlation between the presence of planets and reduced stellar lithium content. The original analysis in \citetalias{hebb2010} utilised only 34 spectra from the CORALIE spectrograph \citep{queloz2000}, but thanks to an ongoing radial velocity observation program we now have access to additional data on WASP-19\,A. Since the publication of \citetalias{hebb2010} a further 3 spectra have been taken using CORALIE, and 36 spectral measurements have been obtained using the HARPS high precision echelle spectrograph \citep{mayor2003}. We co-added the individual spectra into a single spectrum with a higher signal-to-noise ratio, from which we were able to improve the constraint on the lithium abundance. Hints of a lithium line were present but only at the level of the spectral noise, limiting us to an upper limit of $\log(\epsilon_{Li})<0.5$. This allows us to place a lower limit on the age of WASP-19\,A of $2.0$\,Gyr \citep{sestito2005}, making an older age for the star more palatable.

The final piece of evidence to be considered is the proximity of WASP-19,\b to the Roche limit. Using the observed system parameters and their associated uncertainties with the formulation of \citet{eggleton1983}, we calculate that the Roche limit for WASP-19\,b is at $0.0148\pm0.0025$\,AU, in agreement with the observed orbital separation of $0.0164^{+0.005}_{-0.0006}$\,AU. \citet{hellier2011} place the planet slightly further from the star at 1.21 times the Roche tidal radius, but we note that they have used a different formulation for calculating the limit. 

We conclude that the evidence is in favour of WASP-19\,A being old. Whilst there is evidence for the star being young, it is more circumstantial than that which points to an older star. The upper limit on the lithium abundance and the results from stellar model fitting in particular point towards an age in excess of $1.0$\,Gyr. If WASP-19\,A is indeed old, then the exploration of tide-governed evolution presented herein suggests that WASP-19\,b has spun up its host star, and might be in the final stages of spiralling into the Roche limit. During our exploration of possible evolutionary histories we found that those which returned a stellar age of $>1.0$\,Gyr tended to exhibit a very short remaining lifetime. There were, however, some histories that married an older age for the star to a long remaining lifetime for the planet, so we are unable to completely rule out that scenario.

\section{Summary and conclusions}
\label{sec:conclude}
We have calculated an age for the WASP-18 system of $0.579^{+0.305}_{-0.250}$\,Gyr, in agreement with the $0.630^{+0.950}_{-0.530}$\,Gyr age found from stellar isochrones by \citetalias{hellier2009}. Using an MCMC algorithm we find tidal quality factors of $\log(Q'_s)=8.21^{+0.90}_{-0.52}$ and $\log(Q'_p)=7.77^{+1.54}_{-1.25}$. Our results imply that WASP-18\,b will reach the Roche limit in $0.076^{+0.790}_{-0.044}$\,Gyr, and that in most cases it will cause its host to spin up as it does so. We are unable to place stronger constraints the status of the system with respect to planetary infall owing to the range of evolutionary histories that we find to be compatible with its observed parameters, but a large number of the evolutionary tracks investigated imply that the planet is in the process of spiralling in to its host star. Our results for the WASP-18 system seem roughly consistent with the work of \citet{matsumura2010}.

For the WASP-19 system we found tidal quality factors of $\log(Q'_s)=6.47^{+0.67}_{-0.95}$ and $\log(Q'_p)=6.75^{+1.86}_{-1.77}$. These values give a stellar age of $1.60^{+2.84}_{-0.79}$\,Gyr, in broad agreement with the constraint of $\mbox{age}\geq1.0$\,Gyr found by \citetalias{hebb2010}, and imply a remaining lifetime of $0.0067^{+1.1073}_{-0.0061}$\,Gyr. We have investigated the possibility that WASP-19\,A is younger than previously estimated, in line with the predictions of gyrochronology. After considering the evidence for both the old and young stellar age possibilities, we conclude that the older age is more probable based on a reanalysis of the spectral lithium abundance and updated stellar model fits. We therefore suggest that WASP-19\,b could be in the final stages of its spiral-in, and could be on the verge of reaching the Roche limit. We found that this scenario was more prevalent amongst those evolutionary histories with a stellar age $>1.0$\,Gyr, but that there were some instances in which the older stellar age coincided with a substantial remaining planetary lifetime. We are therefore unable to rule out the possibility that it will be some time before the planet falls into the star.

Tidal interactions between these two hot Jupiters and their host stars will dramatically affect the evolution of the stellar rotation periods, counteracting and then reversing the spin-down that is expected from evolution according to a Skumanich-type magnetic braking law. The observed rotation periods are irreconcilable with such an evolution of the rotation period, and strongly suggest that falling in hot Jupiters cause their host stars to spin up during their inward, tidal interaction governed migration. It is worth noting that the our results seem to point toward a more diverse range of stellar tidal dissipation strength than is commonly considered in the literature. The ranges of $\log(Q'_s)$ that we have attributed to the two systems investigated herein are slightly disparate, for which there may be several possible explanations. We turn to the work of \citet{pinsonneault2001}, which shows that the mass of the convective zone is a function of $T_{eff}$, and therefore of spectral type. From their fig.\,1, we note that the effective temperature of WASP-18\,A lies close to the point at which the mass of the convective zone becomes negligible, implying a convective zone mass of $M_{CZ}\approx0.001$\,$\mbox{M}_{\mbox{sun}}$. WASP-19\,A, with its much lower effective temperature, would have $M_{CZ}\approx0.030$\,$\mbox{M}_{\mbox{sun}}$. In our work we have assumed that the star rotates as a single body, but we have made no assumption about where the majority of tidal dissipation takes place. If this process occurs in the convective zone then the discrepancy in the masses, and therefore depths, of the convective zones of the two stars could provide an explanation for the disagreement as to the value of $Q'_s$, with a larger convective zone allowing for more efficient dissipation and hence a smaller quality factor.

This study also provides a warning against using gyrochronology to estimate the ages of hot Jupiter host stars. Owing to the tidal spin-up of its host star by the in-falling planet, the age that we have found for the WASP-19 planetary system is greater than the age found using gyrochronology alone. The situation for the WASP-18 system is less clear cut, but there is no doubt that gyrochronology will not provide an accurate estimate of the system age at all points during its evolution. We therefore suggest that care should be taken when applying gyrochronology to hot Jupiter systems. The two systems studied herein are extreme examples of this class of planet, with extremely short orbital periods and very close orbits. It is likely that there will be a critical point in semi-major axis space, beyond which tidal interactions will be sufficiently weak so as not to influence the stellar rotation period. For systems in this region gyrochronology may well work very well. But for systems such as WASP-18 and WASP-19 which have $a<a_{crit}$, gyrochronology should be applied with care. Determination of this critical point will require further work, and it is likely to depend on a host of factors.

\citet{lanza2010} also suggests that gyrochronology may not always provide accurate age estimates for exoplanetary systems, finding that plotting $P_{rot}t^{-\alpha}$ as a function of $T_{eff}$ for planet hosting stars gives a poor fit to the period-colour relation of \citet{barnes2007}. This implies that exoplanet host stars are systematically faster rotators than stars with similar ages and properties that do not appear to have any associated planets. \citeauthor{lanza2010} also finds that the rotation period evolution of F- and G-type planet hosts does not differ substantially from similar stars without hot Jupiters, a conclusion with which we disagree although we note that the final spiral-in of the planet does not appear to have been considered. 

The substantially reduced rotation period that results from tidal spin-up may provide a means of detecting stars that have either been planet hosts in the past, or that have unseen planetary companions that are in the process of spiraling in to the Roche limit. Measurement of such an anomalous rotation period would provide a strong indication of the current or previous existence of a hot Jupiter around that star. Searching for such systems could help to pinpoint targets for extra-solar planet searches.

\section*{Acknowledgments}
\label{sec:acknowledge}
DJAB would like to thank Keith Horne and Moira Jardine for useful comments and suggestions made during an early draft of this manuscript. The authors are grateful to the referee Brad Hansen, for insightful and thought-provoking comments that greatly improved the quality of the manuscript. The WASP Consortium consists of representatives from the Universities of Keele, Leicester, The Open University, Queens University Belfast and St Andrews, along with the Isaac Newton Group (La Palma) and the Instituto de Astrofisca de Canarias (Tenerife). The SuperWASP and WASP-S Cameras were constructed and operated with funds made available from Consortium Universities and PPARC/STFC. This research has made use of NASA's Astrophysics Data System Bibliographic Services, the ArXiV preprint service hosted by Cornell University, and the SIMBAD database, operated at CDS, Strasbourg, France.

\bibliographystyle{mn2e}

\bsp

\label{lastpage}

\end{document}